\documentclass[12pt,preprint,a4paper]{aastex}

\usepackage{graphics,epsf}
\usepackage{amsmath}                % American Mathematical Society package
\usepackage{amsfonts}               % American Mathematical Society fonts
\usepackage{amssymb}                % American Mathematical Society symbol
\usepackage{epsfig}                 % EPS figures
\usepackage{appendix}

\def \K{~\rm{K}}

\def \AU{~\rm{AU}}

\def \yr{~\rm{yr}}

\begin{document}

\title{CONNECTING PLANETS AROUND HORIZONTAL BRANCH STARS WITH KNOWN EXOPLANETS}

\author{Ealeal Bear\altaffilmark{1} and Noam Soker\altaffilmark{1}}

\altaffiltext{1}{Department of Physics, Technion$-$Israel
Institute of Technology, Haifa 32000 Israel;
ealealbh@gmail.com; soker@physics.technion.ac.il.}
\setlength{\columnsep}{1cm} \small

\begin{abstract}
We study the distribution of exoplanets around main sequence (MS)
stars and apply our results to the binary model for the formation
of extreme horizontal branch (EHB; sdO; sdB; hot subdwarfs) stars.
{{{By Binary model we refer both to stellar and substellar
companions that enhance the mass loss rate, {{{{where substellar
companions stand for both massive planets and brown dwarfs.}}}}
}}} We conclude that sdB (EHB) stars are prime targets for planet
searches. We reach this conclusion by noticing that the bimodal
distribution of planets around stars with respect to the parameter
$M_p a^2$, is most prominent for stars in the mass range $1M_\odot
\la M_{\rm star} \la 1.5M_\odot$; $a$ is the orbital separation,
$M_{\rm star}$ is the stellar mass and $M_p$ the planet mass. This
is also the mass range of the progenitors of EHB stars that are
formed through the interaction of their progenitors with planets
(assuming the EHB formation mechanism is the binary model). In the
binary model for the formation of EHB stars interaction with a
binary companion or a substellar object (a planet or a brown
dwarf), causes the progenitor to lose most of its envelope mass
during its red giant branch (RGB) phase. As a result of that the
descendant HB star is hot, i.e., an EHB (sdB) star. The bimodal
distribution suggests that even if the close-in planet that formed
the EHB star did not survive its RGB common envelope evolution,
one planet or more might survive at  $a \ga 1 \AU$. Also, if a
planet or more are observed at $a \ga 1 \AU$, it is possible that
a closer massive planet did survive the common envelope phase, and
it is orbiting the EHB with an orbital period of hours to days.
\end{abstract}

%\keywords{subdwarfs � binaries}

% ==========================================================
\section{INTRODUCTION}
\label{sec:intro}
% ==========================================================

Horizontal branch stars (HB) are Helium burning stars that have
evolved from main sequence stars (MS) through the red giant branch
(RGB). During the RGB phase the star loses a non-negligible amount
of mass. The amount of mass lost determines the properties of the
descendant HB star; namely, its location on the HR diagram. The
distribution of HB stars on the HR diagram, called HB morphology,
has become a growing field of research because HB stars can be the
main UV radiation source in old population (Dorman et al. 1993;
Bertelli et al. 1996), their formation contains some unsolved
problems, and HB stars can even act as standard candles (Fusi
Pecci et al. 1996b). The formation and evolution of HB stars
depend on the mass of the progenitor on the main sequence, initial
Helium abundance (D'Antona et al. 2002), metallicity, and what is
more relevant to our study, the mass lost during the RGB phase
(Fusi Pecci et al. 1996a; Dorman et al. 1995).

HB stars with low mass envelopes have small radii and they are hot.
They are called extreme HB (EHB) stars {{{{ in photometric classification}}}}
(other names are sdO or sdB or hot subdwarfs
{{{{ according to spectroscopic classification; }}}}
in this work we will use all these terms indistinguishably).
To become an EHB star, the RGB progenitor must
lose most of its envelope. The reason for some RGB stars to lose
so much mass is a major unsolved issue in stellar evolution.
The debate is whether a single star (e.g., Yi 2008) can account for
the formation of hot subdwarfs, or whether binary evolution is
behind the hot subdwarf phenomenon (e.g., Han et al. 2007).
Supporting the binary model is the finding that about half of the
sdB stars in the field (not in globular clusters) reside
in close binaries with periods as short as one day or less (Maxted
et al. 2001; Napiwotzki et al. 2004); the companions are either
low-mass main sequence stars or white dwarfs {{{{ (WDs; e.g., Han et al. 2003;
 Geier et al. 2010, and references therein).}}}}
Because the components' separation in these systems is much less than the size
of the subdwarfs' RGB progenitors, these systems must have
experienced a common envelope (CE) phase (e.g., Han et al. 2002,
2003), where the lower mass companion spirals inside the bloated
envelope of the RGB star and finally ejects it.

{{{{ Most of the formation channels of sdB stars are summarized by Han et al. (2003), although
they omit the substellar channel. Stellar binary interaction can result in a stable Roche lobe overflow
(RLOF). In that case an sdB star is formed, but the orbital separation stays large.
A different scenario discussed in the literature is the merger of two helium WDs, as suggested
by Webbink (1984; for more recent papers with more references see
Han et al. 2003, Heber 2008 and Nelemans 2010). In this scenario, gravitational wave radiation
causes two WDs with a small orbital separation to coalesce and form an sdB star.
This scenario is supported by the research of sdB and sdO mass range done by Zhang et al. (2010).
A very recent population synthesis of binary stars is reported by Nelemans (2010).
His conclusion is that both interaction of RGB stars with substellar companions, and merger of He WDs
can contribute to the formation of sdB stars and single He WD.
However, the large number of single He WDs suggest that most of them are the descendent of the interaction
of RGB stars with substellar companions.  The population synthesis of Nelemans (2010) shows
that different aspects of the interaction of RGB stars with substellar objects must be studied.
Our present paper aim at comparison with known exoplanet properties. }}}}

The CE ejection channel provides a reasonable explanation for the
extra mass loss required to form sdB stars. But for about half of
all analyzed subdwarfs there is no evidence for close stellar
companions. {{{Moreover, in globular clusters stellar companions
cannot explain the formation of sdB stars (Catelan 2009). }}} A
solution which has received a major boost with the recent
discovery by Geier et al. (2009), is that substellar objects
influence the evolution of the RGB progenitor (Soker 1998a;
Nelemans \& Tauris 1998; Soker \& Harpaz 2000, 2007; Soker \&
Hershenhorn 2007; Politano et al. 2008; {{{ Villaver \& Livio
2007, 2009; Carlberg et al. 2009; }}} Bear \& Soker 2010; {{{ Nordhaus et al. 2010}}}).
{{{{It is also possible that the planets were formed along
with the sdB star (second generation planets, see Perets 2010) and are a result of the
merger of the two WDs (Silvoti 2008). This explanation has the same shortcomings discussed before.}}}}
{{{{ In the present paper substellar objects will stand for both massive planets and brown dwarfs.
Grouping of brown dwarfs and massive planets (gas giant planets) has its own merit.
Lovis \& Mayor (2007), for example, raise the possibility that massive planets and brown dwarfs
are formed in the same process.
A key issue here is that it is more easy to detect brown dwarfs, and they are more likely to survive
the RGB phase.
However, as there are more planets than brown dwarfs, they are likely to play a larger role
than brown dwarfs. Also, as the statistics for planets around main sequence stars is much better than
that for brown dwarfs, in this paper we deal only with planets.
Adding brown dwarfs will further increase the merit of the planet-induced formation of
sdB stars. }}}}

{{{{ Substellar objects are known to accompany many different
stars in different stages of their life. There is a large body of
literature and research on relevant substellar companions. Here we
mention a few examples. Machalek et al. (2010) study XO-3b, a high
mass hot Jupiter planet ($M_p=11.79\pm 0.59M_J$), on the verge of
deuterium burning, and orbiting an F5V parent star. Another
example is the detection of substellar companion with a mass of
$M_2 \sin(i)=2.9M_J$ that orbits HD145457 (a K0 giant of
$1.9M_\odot$) with an orbital period of $P=176$d (Sato et al.
2010). As indicated by Schuh et al. (2010), the increasing number
of substellar companions to sdB stars may indicate the existence
of an undiscovered population.
 We do not wish to solve the question of how sdBs were formed, as of now it seems that each scenario might
be possible under specific conditions (Geier et al. 2009; Soker 1998a; Han et al. 2002, 2003; Lisker et al. 2005;
Nelemans 2010). We deal here with the substellar scenario (planet induced) for the formation of sdB stars. }}}}

Geier et al. (2009) announced recently the discovery of a close substellar companion to
the hot subdwarf (EHB) star HD 149382.
The orbital period is very short, 2.391~days, implying that the substellar companion had
evolved inside the bloated envelope of the progenitor RGB star (a CE phase).
The mass of the companion is $8-23 M_J$, so either it is a planet
or a low mass brown dwarf.
This discovery supports the prediction of Soker (1998a) that such planets can survive the
common envelope (CE) phase, and more relevant to us, that planets can enhance the mass loss
rate on the RGB and lead to the formation of EHB.
Other planets that orbit EHB at larger separations have been detected
(Silvotti et al. 2007; Lee et al. 2009; Qian et al. 2009).
Silvotti et al. (2007) announced the detection of a planet with a mass of $3.2 M_{\rm J}$,
an orbital separation of $1.7 \AU$, and an orbital period of $P=3.2 \yr$  around the
hot subdwarf V391 Pegasi.
Serendipitous discoveries of two substellar companions around the eclipsing sdB
binary HW~Vir at distances of $3.6 \AU$ and $5.3 \AU$ with orbital periods of
$3321~$d and $5767~$d (Lee et al. 2009) and one brown dwarf around the
similar system HS~0705+6700 with a period of $2610~$d and a separation of $<3.6 \AU$
(Qian et al. 2009) followed recently.
It is quite plausible that closer planets did interact with the RGB
progenitor of the sdB star; they are not observed in these systems.
In the present paper we examine whether the known exoplanets support such a scenario.
We end by noting that all these substellar companions have been detected in the field. It is commonly
assumed that planets don't exist in large enough numbers in globular clusters.
However, one planet has been detected in the M4 globular cluster (Sigursson et al.
2003; Beer et al. 2004 and references there in), and the role of planets in the formation of
EHB in globular clusters, where metallicity is very low, is an open question.
In this paper we are aiming at field stars.

% ==========================================================
\section{THE POSSIBLE ROLE OF PLANETS IN FORMING EXTREME HB STARS }
\label{sec:role}
% ==========================================================
{{{
% ========
\subsection{Relevant processes}
% ========
The main role of planets in enhancing mass loss rate \emph{does not} come from the deposition
of gravitational energy. It is a common practice to take the
gravitational energy deposited by the spiraling-in companion to be
equal to the binding energy of the ejected envelope when
considering stellar companions. For the processes of sdB formation
this was employed by, e.g., Han et al. (2002). Brown dwarfs and
very massive planets (Nelemans \& Tauris 1998) might also expel
most of the envelope by deposition of gravitational energy.
However, as was mentioned in many papers (Soker 1998a, 2001;
Nelemans \& Tauris 1998; Siess \& Livio 1999a,b; Soker \& Harpaz
2000, 2007; Livio \& Soker 2002; Soker \& Hershenhorn 2007;
Politano et al. 2008; Bear \& Soker 2010; Nordhaus et al. 2010; Carlberg et al. 2009),
planets can play a role by imposing different processes than gravitational energy deposition.
The processes that are listed below where planets enhance the mass loss
rate, are less efficient than deposition of gravitational energy
by stellar companions. Therefore, they can play a significant role
only when the primary is a not-too-massive RGB or AGB star. These
stars have a relatively high mass loss rate (even  before the
enhancement by the planet) due mainly to radiation pressure on
dust.

\subsubsection{Spinning-up RGB and AGB Envelopes and Magnetic Activity}
As evident from the title of the a review by Soker 2004
angular momentum deposition plays a crucial role.
In a recent paper Carlberg et al. (2009) found that the known exoplanets are
indeed capable of creating rapid RGB stellar rotators.
A process relevant to rotation is magnetic field amplification
(Soker \& Harpaz 1992; Soker 1998b, 2001; Livio \& Soker 2002; Nordhaus \& Blackman 2006).
We know that the sun possesses a prominent magnetic activity even with a
rotation rate that is less than one per cent of its break up velocity.
RGB and AGB stars have extended convective envelopes, with a relatively fast
convective velocity.
Therefore, in these stars as well, slow rotation might
be sufficient to trigger magnetic activity.
{{{{ Strengthening our point is the detection of a magnetic field of a few kGauss in a few sdB stars
(O'Toole et al. 2005). }}}}
Soker (2000) found that a planet with a mass of $M_p \ga 0.01 M_J$, where $M_J$ is Jupiter mass,
might spin-up the envelope up to $\sim 10^{-4}$ times the break-up velocity.
{{{{ Soker considered a case at the end of the AGB, when the envelope mass can be very low.
As we deal with RGB stars, the planets must be more massive than this limit, i.e., $M_p \ga  M_J$. }}}}
According to Soker (1998b) such an angular velocity might be sufficient to trigger the
formation of magnetic spots, as in the Sun.
Above such cool spots dust forms much more easily, and mass loss rate is increased.
As we will study planets of masses $M_p \ga M_J$, we do expect magnetic activity to
enhance mass loss rate.

{{{{ We can estimate the angular velocity of the RGB star.
We take the  moment of inertia as $I \simeq 0.2 M_{\rm env} R_{\rm env}^2$,
and the planet to enter the envelope, due to tidal interaction, from a distance
of $a \simeq 4 R_{\rm star}$.
From conservation of angular momentum we find the angular velocity after the onset of the CE to be
\begin{equation}
\Omega \equiv \frac{\omega}{\omega_{\rm c}} \simeq 10 \frac{M_p}{M_{\rm star}},
\label{Eq.omega}
\end{equation}
where $\omega_{\rm c}$ is the critical (break-up) angular velocity of the RGB stars.
For an envelope mass of $M_{\rm env} \sim 0.5M_\odot$ and a planet mass of $M_p \sim M_J$,
we find $\Omega \simeq 0.02$. This is non-negligible if we recall that the sun has $\Omega \simeq 0.005$
and posses a clear magnetic activity. }}}}

\subsubsection{Excitation of Waves in Common Envelopes}

One such nonlinear effect is the excitation of p-waves
(Soker 1992, 1993) during a common envelope phase.
While inside the convective envelope of an AGB or RGB star,
a companion will excite p-waves which propagate outward with
increasing amplitude, mainly in the equatorial plane.
The surface, $r=R$, relative pressure amplitude $P^\prime/P$ on the
equatorial plane is given by
\begin{equation}
\left(\frac{\vert P^\prime \vert}{P} \right)_{r=R}
\simeq 0.03 \frac{M_p}{M_J} \left(\frac{a_2}{0.1R} \right)^{-0.75},
\label{eq:pressure}
\end{equation}
where $a_2$ is the orbital separation between the planet and the core.
The weak dependance on the mass of the primary star, and
the small damping of the propagating waves in the convective envelope were
averaged for typical numbers (see Soker 1993 for details).
The perturbation increases linearly with the companion
mass (for low mass companions), and increases somewhat
as $a_2$ decreases (depending on convective viscosity).
The amplitude is much larger in the equatorial plane than in the
polar directions.
Such excited non-radial oscillation can enhance mass loss rate in the equatorial plane.

{{{{ It is important to note that the extra energy carried by
the stronger wind does not come from the energy carried by the
excited waves. The energy comes from the RGB radiation. The
planets cannot supply the required energy in waves. The waves only
perturbed the surface, enhancing mass loss rate, e.g., by
facilitating dust formation. This process demands further study,
which is beyond the scope of the present paper.   }}}}

\subsubsection{Destruction of Planets in the Envelope}

A study of the fate of planets in the envelope of AGB stars
was conducted by Livio \& Soker (1984). They assumed that the
planet accretes from the envelope at the Bondi-Hoyle accretion rate.
However, it is possible that the planet swells as a
result of this accretion and does not accrete much,
like low-mass main sequence stars do (Hjellming \& Taam 1991).
Planets may also be evaporated, in particular when they
reach the place in the envelope where the envelope's temperature
exceeds the planet's virial temperature.
For stars on the upper RGB and AGB, the orbital separation of a planet
from the core where fast evaporation starts is (Soker 1998a)
\begin{equation}
a_2({\rm evaporation}) \simeq 10
\left( \frac{M_p}{M_J} \right)^{-1} R_\odot.
\end{equation}
The cool and dense evaporated material is still of low entropy,
and fraction of it may spiral-in to the core. More massive planets
than Jupiter will survive farther in, until they reach a radius
where Roche lobe overflow (RLOF) occurs. For a planet of radius
$R_p=0.1 \eta R_\odot$ {{{{ (this equality defines $\eta$ as
the ratio of the planet radius to $0.1 R_\odot$), }}}} RLOF occurs
when the orbital separation from the core is (see Soker 1998a)
\begin{equation}
a_2({\rm RLOF}) \simeq 1.7 \eta
\left( \frac{M_p}{M_J} \right)^{-1/3} R_\odot.
\end{equation}
The addition of the disrupted planet (or brown dwarf) material to
the core and around it may have several effects. First, the low
entropy material can absorb heat, and may reduce for a short
period of time the stellar luminosity (Harpaz \& Soker 1994).
Second, if the material reaches the core, or close to it, the
release of gravitational energy and nuclear burning of the fresh
hydrogen-rich material may lead to stellar expansion and enhanced
mass loss rate (Siess \& Livio 1999a,b). Third, the high specific
angular momentum of the planet's (or brown dwarf) material may
lead to the formation of an accretion disk around the core; such
disk can launch two jets (Soker 1996).

% ========
\subsection{The relevant parameter}
% ========

If there were enough planets, we would conduct our analysis in the $M_p-a$ plane
(planet's mass$-$orbital separation plane ). Since the number of planets
relevant to our analysis is limited (as we show in the next sections), we seek one
parameter of the form of $M_p a^{\beta}$.

As discussed above, angular momentum is a parameter more relevant than
gravitational energy deposition.
The primary stellar mass of our study is in a relatively narrow range of $\sim 1-1.5 M_\odot$.
Therefore, the angular momentum of planet is proportional to $M_p a^{1/2}$, where $a$ is
the initial orbital separation.
However, a planet will influence the mass loss process much more if it is engulfed when the
primary is a larger RGB star. The mass loss rate of an isolated RGB star is a strong
function of the stellar radius. For example, for a constant RGB effective temperature
$(\sim 3500 \K$), the Reimers mass loss rate from the RGB star varies as $\sim LR/M \sim R^3$.
Overall, if we seek a parameter of the form $M_p a^{\beta}$, it should be with $\beta \gg0.5$.
As we have no a priori value to use, we follow Soker \& Hershenhorn (2007) and take $\beta=2$.
This value is not fundamental, but it captures the essence of the processes discussed above,
with the sensitivity to the planet angular momentum and the requirement that the
star be a well developed giant for the planet to influence its mass loss.
We expect that in about a decade the number of exoplanets will alow a much deeper study.
}}}
% ==========================================================
\section{RELEVANT PROPERTIES OF EXOPLANETS}
\label{sec:proprties}
% ==========================================================

More than 400 exoplanets are known, detected through the use of
several different techniques (e.g., Bennet 2009; Crouzet et al.
2009; Gregory 2009; Beckwith 2008; Hebrard et al. 2009, and many
more references within these recent papers; see the The Extrasolar
Planets Encyclopedia edited by Jean Schneider;
http://exoplanet.eu/). These are prone to different selection
effects that influence any statistical analysis. Many statistical
analyses have been done since the first exoplanet discovered:
e.g., an extensive analysis of the eccentricity and the
architecture of multiplanet systems (Marcy et al. 2005). The
general consensus is that the existence of planets rises with
metallicity (e.g., Greaves et al. 2007; Soker \& Hershenhorn 2007;
Santos et al. 2003). Many studies have tried to find a prominent
connection between the orbital parameters, the star mass, and its
metallicity (e.g. Mushaliov et al. 2009; Wright et al. 2009;
Santos 2008; Cumming et al. 2008; Desidera \& Barbieri 2007;
Mugrauer et al. 2007; Ribas \& Miralda-Escude 2007; Udry \& Santos
2007; Ida \& Lin 2005; Halbwachs et al. 2005; Rice \& Armitage
2005; Santos et al. 2005; Ksanfomality 2004; Udry et al. 2003,
2004; Israelien et al. 2004; Udry \& Mayor 2001).

Despite the above limitations, it seems that there are enough planets to conduct an analysis of the implications of exoplanets
properties for the role they can play in late stellar evolution
(beyond the main sequence). As we are interested in the influence of planets on
RGB stars, we follow  Soker \& Hershenhorn (2007) that had a similar goal.
Soker \& Hershenhorn (2007) examined the number of planets as a function of metallicity bins
and the planet mass $M_p$, orbital separation $a$, and orbital eccentricity $e$, in several combinations.
They found that planets orbiting high metallicity stars
tend to part into two groups in a more distinct way than planets orbiting low metallicity stars.
They also found that high metallicity systems tend to produce planets which reside
in closer orbital period separation on average.
Soker \& Hershenhorn (2007) had 207 planets in their analysis. We repeated their analysis using 331 planets
(out of more than 400 that were discovered so far) and got similar results.

The progenitors of EHB stars in the field are MS stars in the mass range $1M_\odot \la M \la 1.5M_\odot$,
which we term here the {\it middle range}.
{{{{ The connecting link between MS star with planets and EHB with planets
is the RGB phase. Although it is likely that close, low mass planets will be engulfed during this stage, some planets might
escape this scenario. Villaver \& Livio (2009) note that until today about 20 exoplanets have been discovered
around giant stars. Among them are the massive planets of minimum mass of $10.6M_J$ and $19.8M_J$ at the
open clusters of NGC 2423 and NGC 4349, respectively.
These planets have large orbital periods of 714 and 678 days, respectively (for details see Lovis \& Mayor 2007).
Closer planets might be engulfed in the future or overcome tidal interaction depending on specific
parameters. Other scenarios for the formation of planets around giant stars exist. For example Wickramasinghe et al.
(2010) suggest that a planet surrounding a giant star can form in a rare merger of two WDs. }}}}

\begin{figure}
\includegraphics[scale=0.5]{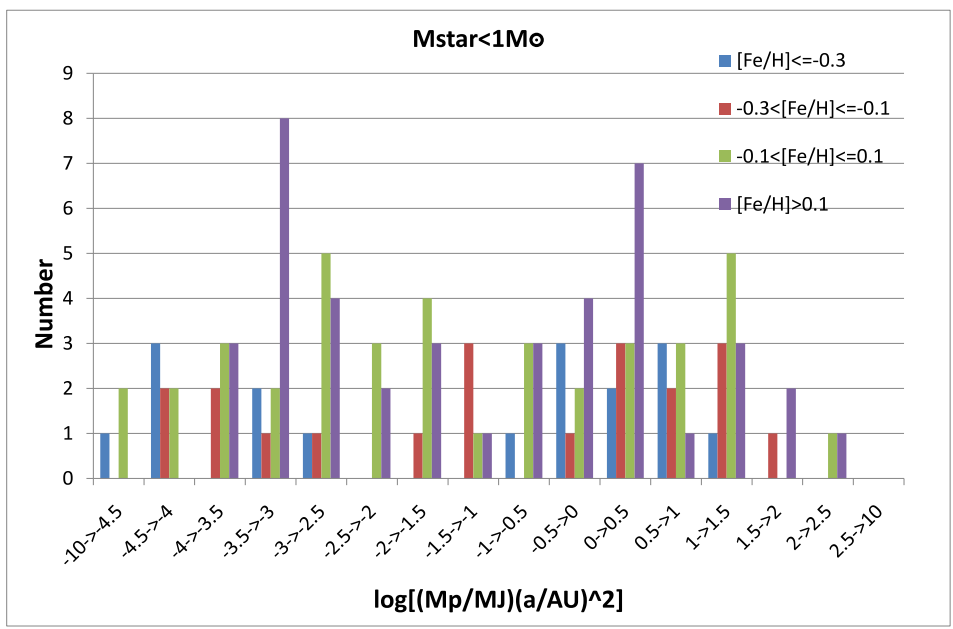}
\caption{Number of planets vs.
%$log[\frac{M_p}{M_J}{(\frac{a}{AU})}^2]$.
$\log \left[ \frac {M_p}{M_J}{(\frac{a}{AU})}^2 \right]$, where $M_p$ is the
planet mass, and $a$ is the orbital separation. The planets
shown in this graph are planets that orbit a star of
$M_{\rm star}< 1M_\odot$. Total of 118 planets.} \label{fig:All_planets_1}
 \end{figure}

\begin{figure}
\includegraphics[scale=0.5]{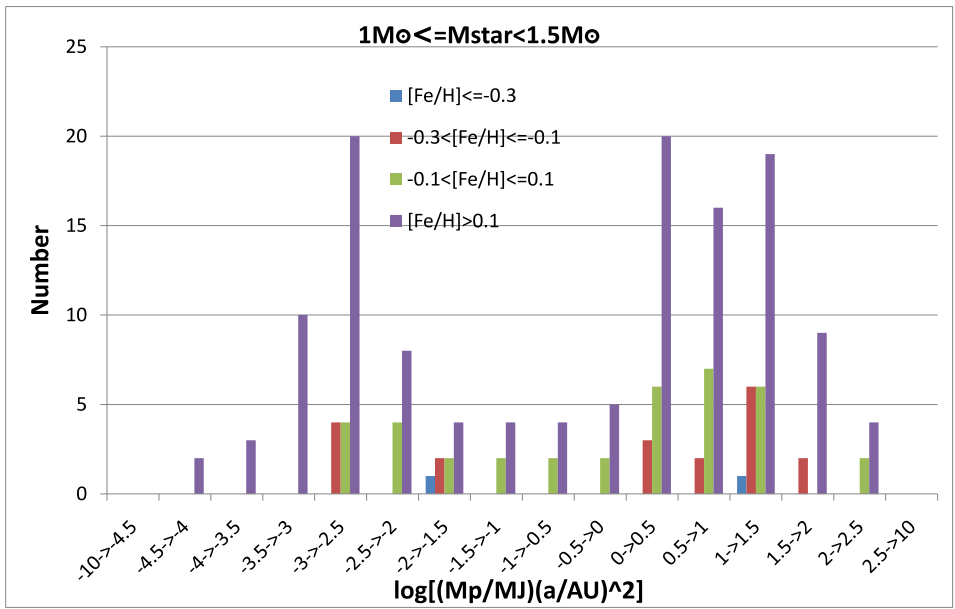}
\caption{Number of planets vs
$\log \left[ \frac {M_p}{M_J}{(\frac{a}{AU})}^2 \right]$. The planets shown in
this graph are planets that orbit a star in
the mass range of $1M_\odot \leq M_{\rm star}<1.5M_\odot$, i.e. middle stellar mass
range. Total of 186 planets. }
\label{fig:All_planets_2}
 \end{figure}

\begin{figure}
\includegraphics[scale=0.5]{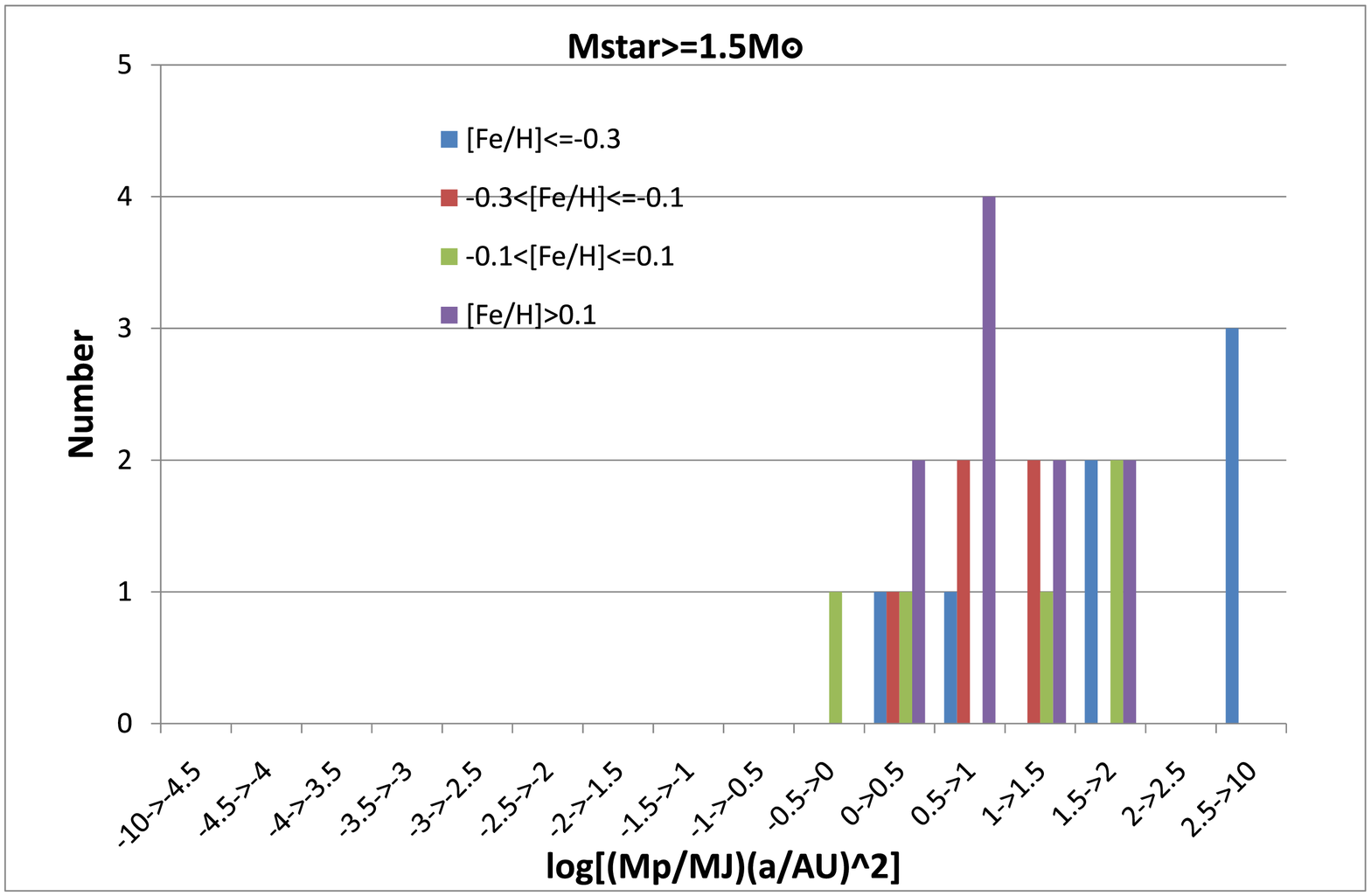}
\caption{Number of planets vs.
$\log \left[ \frac {M_p}{M_J}{(\frac{a}{AU})}^2 \right]$. The planets shown in
this graph are planets that orbit a star in the mass range of $M_{\rm star}\geq 1.5M_{\odot}$. Total of 27 planets.
 We see no bimodality. However, this might be in part due to a selection effect (see text).
}
\label{fig:All_planets_3}
 \end{figure}

The lower limit on the mass of the EHB progenitor comes from the constraint on the evolution time
scale from stellar formation to the HB, while the upper limit comes from the requirement on
the progenitor to lose most of its envelope on the RGB.
We will therefore use the $1M_\odot$ and $1.5 M_\odot$ mass boundaries to divide the stars in our analysis.
Building on the results of Soker \& Hershenhorn (2007), we take $M_pa^2$ to be the main parameter to describe the planets.
This parameter was chosen after trying different parameters, including the orbital separation alone.
We choose to work with $M_p a^2$ rather than $a$ for several reasons (Soker \& Hershenhorn 2007):
(1) The bimodal distribution is smoother.
(2) The gap between peaks is wider (in a logarithmic scale).
(3) There is a relatively clear difference between low and high metallicities. In the low metallicity the distribution is unclear while in the
high end of the metallicity the bimodal distribution is prominent.
(4) There are relatively small fluctuations within the peaks themselves.
{{{Our results are presented in Figs. \ref{fig:All_planets_1} - \ref{fig:All_planets_3} which illustrate the number of
planets vs. the chosen parameter (see Section \ref{sec:role}) for different stellar mass.}}}
We also examine the influence of metallicity, as this is a significant parameter determining planet formation.
The data is taken from the Extrasolar Planet Encyclopedia update to March 03rd 2010.
% FFFFFFFFFFFFFFFFFFFFFFFFFFFFFFFFFFFFFFFFFFFFFFFF
% FFFFFFFFFFFFFFFFFFFFFFFFFFFFFFFFFFFFFFFFFFFFFFFF
% FFFFFFFFFFFFFFFFFFFFFFFFFFFFFFFFFFFFFFFFFFFFFFFF
% FFFFFFFFFFFFFFFFFFFFFFFFFFFFFFFFFFFFFFFFFFFFFFFF
% FFFFFFFFFFFFFFFFFFFFFFFFFFFFFFFFFFFFFFFFFFFFFFFF
% FFFFFFFFFFFFFFFFFFFFFFFFFFFFFFFFFFFFFFFFFFFFFFFF

The bimodal distribution that is best seen in Fig.
\ref{fig:All_planets_2} can teach us about the fate of the planet
during the RGB phase of the parent star. Since the planet masses
are $ M_p > 0.006 M_J$, all stars in the left group will be
engulfed by the RGB star, while some planets in the right group
will avoid the CE phase. Those that avoid the CE phase will move
out because of the mass loss process, and can be detected as
planets orbiting HB stars with orbital separations of several AU
and more. The engulfed planets will increase the total mass lost
by the RGB star, and by that form blue HB stars, including EHB
stars. Massive planets of $M_p \ga 10 M_J$ might survive the CE
phase (Soker 1998a), and later be detected as planets orbiting EHB
star at very short orbital periods (down to several hours). This
general bimodality in survival routes has implications for
multiplanet systems, that we study in section \ref{sec:multi}.

Fig. \ref{fig:All_planets_3} represents planets around massive
stars ($M_{\rm star}\geq 1.5M_{\odot}$). The stars in the figure
belong to diverse groups of giant, subgiant, and MS stars. The
typical radius for these stars is larger than $4R_\odot$, although
some are MS stars, such as $Fomalhaut$ which is an A3V star with
$M_{\rm star}=2.06M_\odot$ and $R_{\rm star}= 1.82R_\odot$. An
example for an evolved star is HD 102272 which is a K0 star of
mass $M_{\rm star}=1.9M_\odot$ and radius $R_{\rm star}=
10.1R_\odot$. This star is unique since although it is relatively
large, it has a multiplanet system around it. The difficulty in
detecting planets around massive stars or evolved stars has been
discussed in the literature (e.g. Hatzes et al. 2005). It is hard
to detect planets around massive MS stars because they are hot.
Therefore, the Doppler shift technique requires them to become
cooler as they evolve toward the RGB. Stars evolving off the MS
are less likely to have a bimodal distribution of planets, since
as their radius increases they are likely to swallow close in
planets (Kunitomo et al. 2010; Carlberg et al. 2010).

 Johnson et al. (2010) claim that the distribution of orbital separations
depend on the mass of the parent star and its evolutionary phase.
Johnson et al. (2010) describe the detection of a close planet
around the subgiant star HD 102956 at an orbital separation of
$a_p=0.081AU$. This is a surprising result since until now
observations show a shortage in close exoplanets ($a_p < 0.6 AU$)
near intermediate stars (Johnson et al. 2010; Burkert \& Ida
2007). This detection is added to other detections, e.g. Dollinger
et al. (2009) who reports the presence of a substellar object with
$M_p\sin i=10.50\pm 2.47$ around 11 UMi ($M_{\rm star}=1.80\pm
0.25M_\odot$). The question if the absence of bimodality around
massive ($M>1.5 M_\odot$) stars is real or a selection effect is
an open question (e.g., Currie et al. 2009; Johnson et al. 2010).

% ==========================================================
\section{Multiplanet systems}
\label{sec:multi}
\begin{figure}
\includegraphics[scale=0.5]{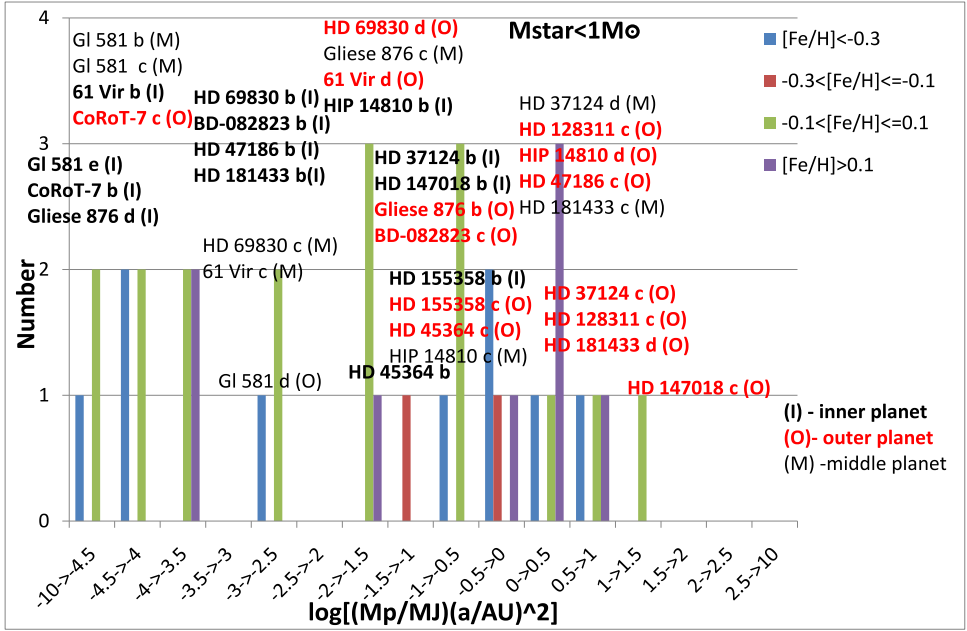}
\caption{Number of planets vs.
$\log \left[ \frac {M_p}{M_J}{(\frac{a}{AU})}^2 \right]$. The planets shown in
this figure are multi planet systems for stars in the mass range of $M_{\rm star}< 1M_\odot$. Where ${\rm (I)}$ signifies
the inner most planet around the star (shortest period/orbital separation);
${\rm (O)}$ signifies the outer most planet (longest period/obital separation);
${\rm (M)}$ refers to the middle planets in case there are more than two planets around the star. Total of 36 planets. } \label{fig:Multi_1}
 \end{figure}

\begin{figure}
\includegraphics[scale=0.5]{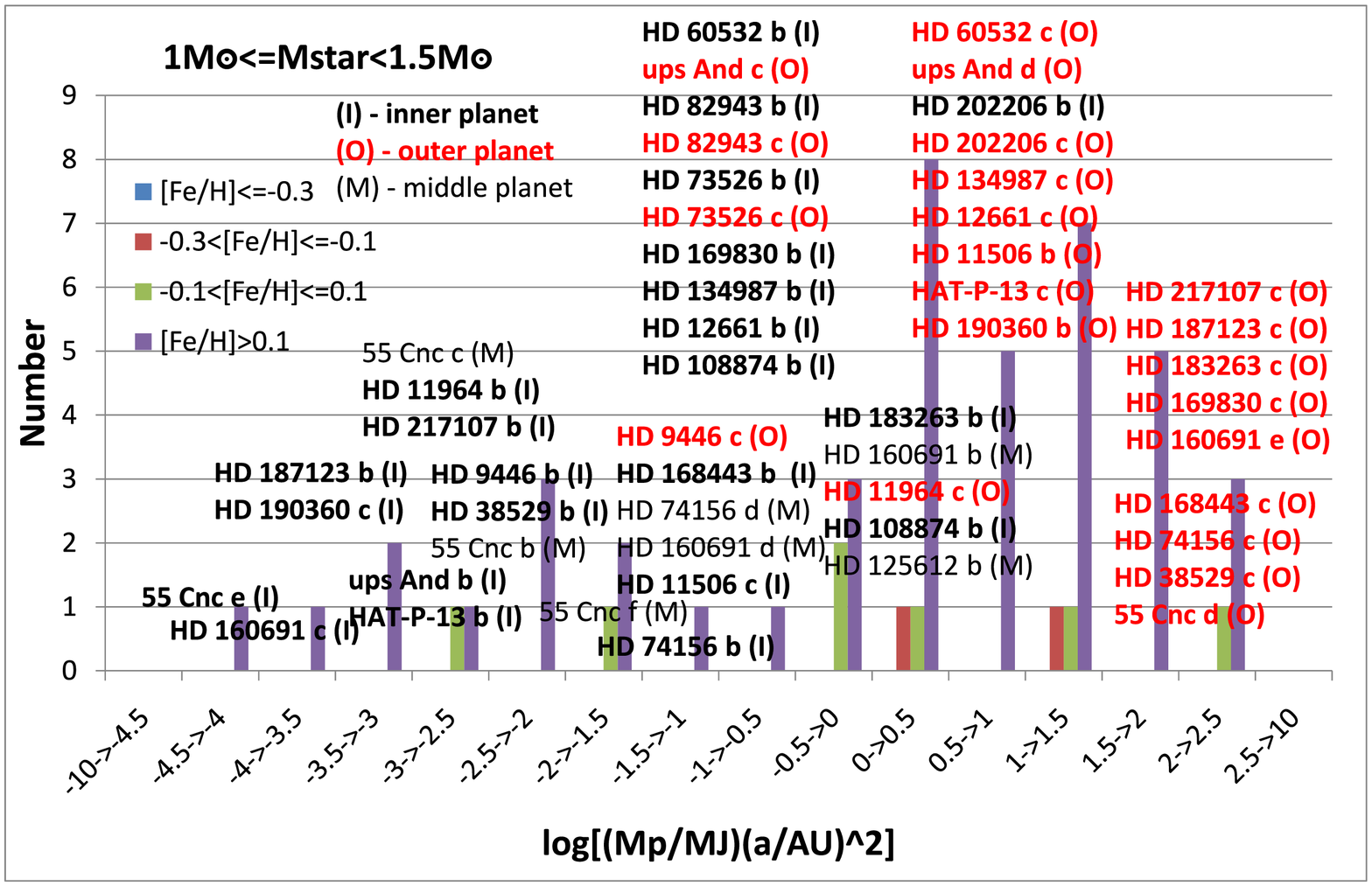}
\caption{Number of planets vs.
$\log \left[ \frac {M_p}{M_J}{(\frac{a}{AU})}^2 \right]$. The planets shown in
this figure are multi planet systems for stars in the mass range of $1M_\odot\leq M_{\rm star}< 1.5M_\odot$ (middle stellar mass
range). Total of 52 planets.} \label{fig:Multi_2}
 \end{figure}

\begin{figure}
\includegraphics[scale=0.5]{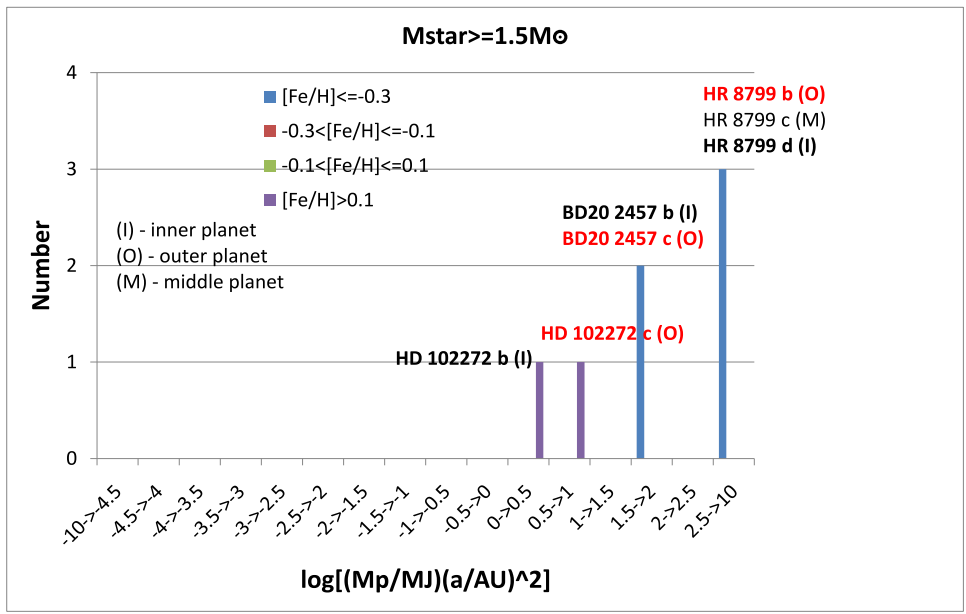}
\caption{Number of planets vs.
$\log \left[ \frac {M_p}{M_J}{(\frac{a}{AU})}^2 \right]$. The planets shown in
this figure are multi planet systems for stars in the mass range of $M_{\rm star} \geq 1.5M_\odot$. Total of 7 planets. } \label{fig:Multi_3}
 \end{figure}

\begin{figure}
\includegraphics[scale=0.5]{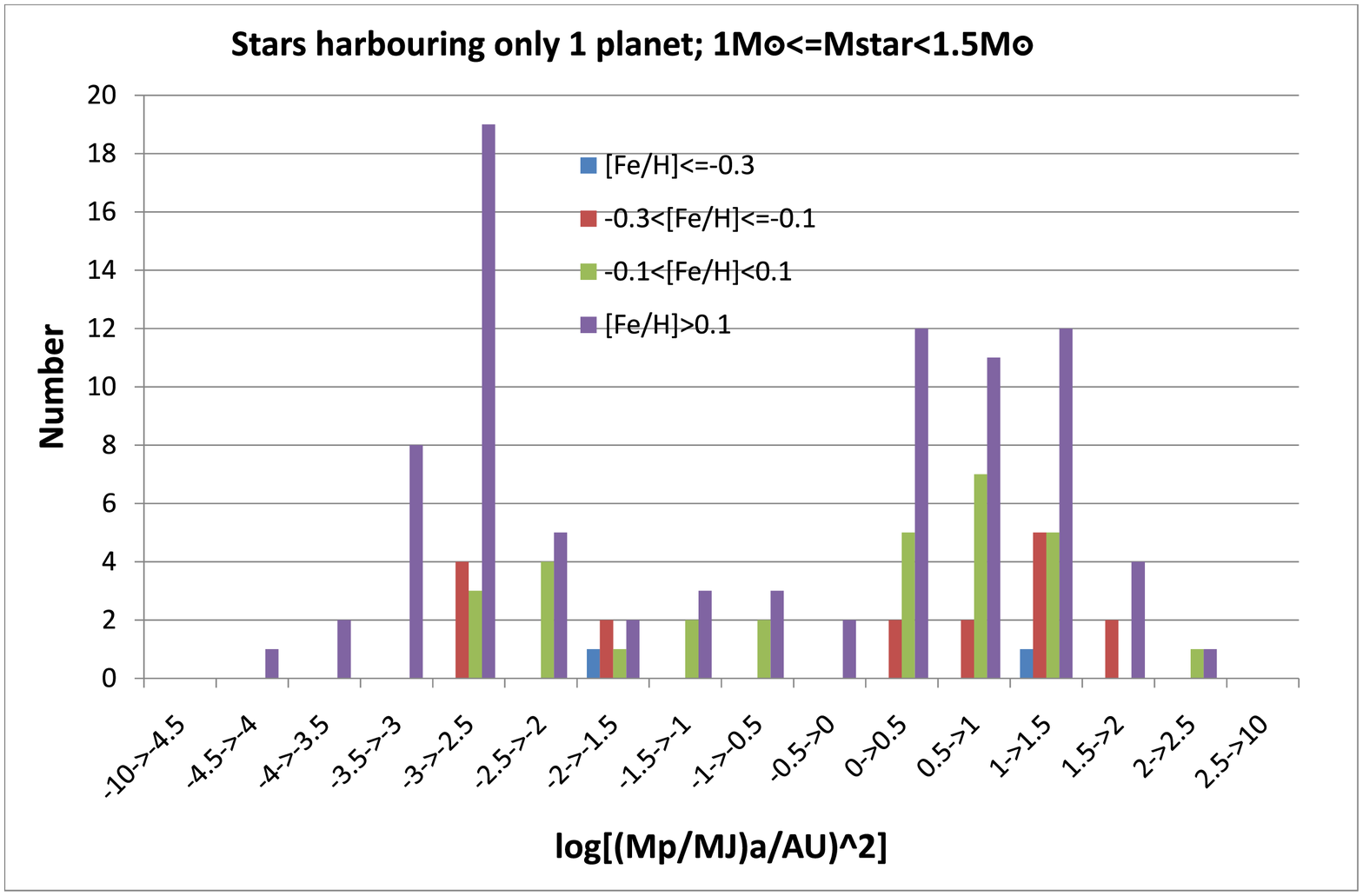}
\caption{Number of planets vs.
$\log \left[ \frac {M_p}{M_J}{(\frac{a}{AU})}^2 \right]$, for stars with only one detected planet for stars in the mass range
of $1M_\odot\leq M_{\rm star}< 1.5M_\odot$.
Total of 134 systems in the figure. } \label{fig:single_2}
 \end{figure}

% ==========================================================
It seems that a significant fraction of planet-host stars have
more than one planet (Wright 2009; Marcy et al. 2005; Udry \&
Santos 2007). We would like to analyze the multiplanet systems
following the results of the previous section, with the goal of
understanding the implications for the formation of EHB (sdO, sdB)
stars. Although migration can alter the position of planets within
a multiplanet system, the migration of planets around EHB is
beyond the scope of this paper. Around MS stars, most of the
migration occurs as a result of interaction with the
proto-planetary disk, which does not exist around EHB stars. We
start by presenting all multiplanet systems around stars from the
Extrasolar Planet Encyclopedia update to March 03rd 2010. This is
done in Figs. \ref{fig:Multi_1}$-$\ref{fig:Multi_3}, for the
stellar mass ranges $M_{\rm star}<1M_\odot$, $1M_\odot\leq M_{\rm
star}< 1.5M_\odot$, and $M_{\rm star}\geq1.5M_\odot$,
respectively, as in the previous section. {{{ These three figures
(\ref{fig:Multi_1}$-$\ref{fig:Multi_3}) represent the number of
planets vs. $M_pa^2$, for different stellar masses. This
representation is similar to the representation in Figs.
\ref{fig:All_planets_1} - \ref{fig:All_planets_3}, except now only
multiplanet systems are presented. Fig. \ref{fig:single_2}
illustrates the number of planets vs. $M_pa^2$ for stars in the
mass range of $1M_\odot\leq M_{\rm star}< 1.5M_\odot$, having only
one known planet. }}} We use the following notations: ${\rm (I)}$
signifies the inner most planet around the star (shortest period/
orbital separation); ${\rm (O)}$ signifies the outer most planet
(longest period/obital separation); ${\rm (M)}$ refers to the
middle planets in case there are more than two planets around the
star. For comparison purposes we present in Fig.
\ref{fig:single_2} the distribution of systems with single
detected planet, in the middle stellar mass range.
% FFFFFFFFFFFFFFFFFFFFFFFFFFFFFFFFFFFFFFFFFFFFFFFF
% FFFFFFFFFFFFFFFFFFFFFFFFFFFFFFFFFFFFFFFFFFFFFFFF
% FFFFFFFFFFFFFFFFFFFFFFFFFFFFFFFFFFFFFFFFFFFFFFFF
% FFFFFFFFFFFFFFFFFFFFFFFFFFFFFFFFFFFFFFFFFFFFFFFF
% FFFFFFFFFFFFFFFFFFFFFFFFFFFFFFFFFFFFFFFFFFFFFFFF
% FFFFFFFFFFFFFFFFFFFFFFFFFFFFFFFFFFFFFFFFFFFFFFFF
% FFFFFFFFFFFFFFFFFFFFFFFFFFFFFFFFFFFFFFFFFFFFFFFF
% FFFFFFFFFFFFFFFFFFFFFFFFFFFFFFFFFFFFFFFFFFFFFFFF

Three main results emerge from Figs. \ref{fig:Multi_1}$-$\ref{fig:single_2}:
\begin{enumerate}
\item In general, there are two groups of planets, in the left and right hand sides of Figs. \ref{fig:All_planets_1}-\ref{fig:single_2}.
The ratio between the numbers of planets in each group strongly depends on the host stellar mass, and to lesser degree on the metallicity.

\item In the middle stellar mass range there is a clear separation to two groups of
planets (left and right sides of Fig. \ref{fig:Multi_2}), mimicking the distribution of all systems, as well
as systems that have only one detected planet (Fig. \ref{fig:single_2}).
There is a specific range in the value of $M_p a^2$ where planets are less likely to
reside in.
\item Stars in the range of $1M_\odot\leq M_{\rm star}< 1.5M_\odot$ (middle stellar mass
range) tend to have high metallicity, but this can also be a selection effect.
\end{enumerate}
{{{ Fig. \ref{fig:Multi_3} depicts a range of massive stars. We will not focus on this range, since these stars
are too massive for planets to cause them to form EHB stars. In addition, the statistics
in that mass range is poor and contains selection effects.}}}

% ==========================================================
\section{Implications for the formation of EHB (sdO, sdB) stars}
\label{sec:implication}
% ==========================================================
The progenitors of EHB stars are MS stars or evolved MS stars that
have not gone through the He flash in the mass range of $\sim 1 -
1.5 M_\odot$ (in globular clusters stars of $\sim 0.9 M_\odot$ are
the progenitors of present HB stars). They are the stars of Figs.
\ref{fig:Multi_2} and \ref{fig:single_2}, that have the strongest
bimodality. Since most planets are in the mass range $0.1-10M_J$,
the bimodality is mainly due to the bimodality in the orbital
separation (that appears as $a^2$ in our parameter $M_p a^2$). The
bimodality in the orbital separation is well known (e.g., Desidera
\& Barbieri 2007; Butler et al. 2006), but it is less prominent
than the bimodality in the parameter $M_p a^2$. According to the
planet formation mechanism of EHB stars, as the MS progenitor
evolves through the RGB phase it engulfs the close planets (Soker
1998a; Nelemans \& Tauris 1998). The planets enhance the mass loss
rate, leading to the formation of a blue-HB star, in many cases
EHB stars.

The bimodality suggests that in many cases there are planets further out
that avoid the CE phase and survive the RGB phase.
If the planet is close, it will come closer due to tidal effects, still avoiding the CE
phase (Bear \& Soker 2010). If it is further out, it will spiral out.
The planet that was engulfed can be either destroyed (Soker 1998a;
Villaver \& Livio 2009; Li et al. 2008; Siess \& Livio 1999a,b), or survive if it is massive enough,
$M_p \ga 10 M_J$ (Soker 1998a; Soker \& Harpaz 2008).

The implications are that many EHB stars that have no stellar companions were formed by
interaction with planets.
Even if the closest planet(s) did not survive, Fig. \ref{fig:Multi_2} suggests that
there might be surviving planets at orbital separations of $a \ga 1 \AU$.
In the opposite direction, detection of planets at $a \ga 1 \AU$ around EHB stars,
strongly suggests that there was indeed a closest planet. There is even a chance
that it survived. Over all, the search for planets around single EHB stars, in all techniques, including transient,
is highly encouraged.

\begin{figure}
\includegraphics[scale=0.5]{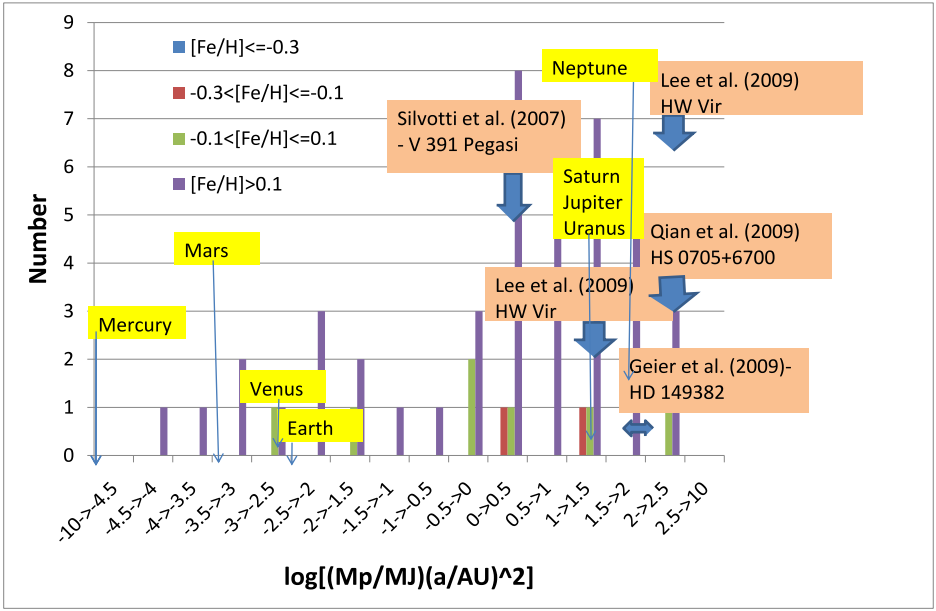}
\caption{Number of planets vs.
$\log \left[ \frac {M_p}{M_J}{(\frac{a}{AU})}^2 \right]$. The planets shown in
this figure are multi planet systems for stars in the mass range of $1M_\odot\leq M_{\rm star}< 1.5M_\odot$, plus the planets of our
solar system. The histogram is the same as in Fig. \ref{fig:Multi_2}.
Systems marked by their names are EHB stars that host planets.
Their location is according to our estimate of the location of the planet
during the main sequence phase of the progenitor.}
\label{fig:Multi_2_var1}
 \end{figure}

Let us apply these conclusions to known substellar objects around HB stars.
As we have no information on the MS stellar masses of the progenitors of these HB stars, we
take the progenitor MS mass of all these systems to be $M_{\rm Pro}=1.25M_\odot$, at the middle of
our middle stellar mass range. Where $M_{\rm Pro}$ is the mass of the progenitor of the EHB.
In a case of an outer planet, that avoided a CE phase, we assume that the orbital separation
increased due to mass loss, and for lack of knowledge we ignore tidal interaction.
We also assume that mass accretion by the planet is negligible.
Therefore, the orbital separation around the MS progenitor was

\begin{equation}
a_0 \simeq \frac{M_{\rm HB}}{M_{\rm Pro}} a_{\rm HB} \qquad{\rm No~CE~phase}
\label{eq:radius}
\end{equation}
where ${M_{\rm HB}}$ is the mass of the present EHB stars, and $a_{\rm HB}$ is the present orbital
separation of the planet. We note that even in the complicated system B1620-26 that
contains a pulsar, a WD, and a planet, mass loss was considered to increase the orbital
separation of the planet (Sigurdsson et al. 2003).
The results of the calculations for the known HB systems listed below are presented in Fig. \ref{fig:Multi_2_var1}.
% FFFFFFFFFFFFFFFFFFFFFFFFFFFFFFFFFFFFFFFFFFFFFFFF
% FFFFFFFFFFFFFFFFFFFFFFFFFFFFFFFFFFFFFFFFFFFFFFFF

(1)\textit{HD 149382.} Geier et al. (2009) discovered a substellar
object surrounding HD 149382 (a hot subdwarf star) with the
following parameters: $M_p= 8 - 23M_J$, $P=2.391d$, $a=5 -
6.1R_\odot$ and $M_{\rm sdB}=0.29-0.53M_{\odot}$ {{{{ (in this
system the inclination is constrained, so that the planet mass can
be considered as known to within the range of the cited
uncertainty.) }}}} As the substellar object experienced a CE
phase, we cannot use equation \ref{eq:radius}. Instead, its
initial separation should be {{{ $a_0 \la 3 R_{\rm RGB}$ (Nordhaus
et al 2010; Villaver \& Livio 2009)}}}. We take for this star
(which is not in the halo) the maximum radius on the RGB to be
$0.5 AU$, hence {{{ $a_0 \la 1.5 \AU$ (Nordhaus et al. 2010)}}}.
Recently a different study by Jacobs et al. 2010, did not find the
planet. Jacobs et al. (2010) analyzed He lines while Geier et al
(2009) analyzed metal lines. Although the existence of this planet
is now questionable (and hopefully will be resolved soon by the
observers listed above), it does not change our conclusions. In
the case it is not a real effect , it only reduces one point from
Fig. \ref{fig:Multi_2_var1}.

(2)\textit{V 391 Pegasi.} Silvotti et al. (2007) reported the
discovery of a planetary mass body orbiting the star V 391 Pegasi
(EHB star) with the following calculated orbital parameters: $a = 1.7\pm 0.1AU$,
$P=3.2 \pm 0.12{\yr}$, $M_p=3.2\pm 0.7M_{J}$ also known as HS~$2201+2610$
{{{{ (the inclination is not constrained, and the planet mass is only a lower limit). }}}}
Following Eq. \ref{eq:radius}, assuming $M_{\rm Pro}=1.25M_{\odot}$ we find that $a_0\simeq 0.68AU$.
By assuming $M_{\rm HB}=0.5\pm0.05M_\odot$, Silvotti et al. (2007) suggest that
in the most likely scenario their detected planet avoided the CE phase, because
the maximum radius at the tip of the RGB is $R_{\rm RGB}=0.7AU$ (Sweigart \& Gross 1978; Han et al. 2002).
Although $a_0>R_{\rm RGB}$, we note that tidal interaction will cause such a planet to spiral in
{{{ from a distance of $a_0 \la 3 R_{\rm RGB}$. Even for $R_{\rm RGB} =0.5AU$ the limited
orbital separation is $a_0 \simeq 1.5AU$ (Nordhaus et al. 2010)}}}.
We conclude that we need to introduce another ingredient to explain how the detected planet avoided the CE phase.
Such can be a closer in planet that enhanced the mass loss rate early on the RGB. The closer in planet
entered a CE and did not survive, but by enhancing the mass loss rate saved the outer planet.

(3)\textit{HS~$0705+6700$.} Qian et al. (2009) reported the discovery of a third member around the binary HS~$0705+6700$,
which is composed of a hot sdB type primary of $M_1=0.483M_\odot$ and a fully convective M-type secondary of
$M_2=0.134M_\odot$; the orbital period is $P=2.3$h (Drechsel et al. 2001).
The third member is a brown dwarf with orbital parameters of:
$P=7.15{\yr}$, $a <3.6AU$, and $M_p\sim 0.072M_{\odot}$, when the total mass for HS~$0705+6700$ is assumed
at $M_{\rm HB}=0.617M_{\odot}$ {{{{ (in this system the
inclination is constrained, therefore the planet mass can be considered as given). }}}}
Following Eq. \ref{eq:radius}, assuming $M_{\rm Pro}=1.38M_{\odot}$ we find that $a_0\simeq 1.4AU$.
We note that in this system the M-type secondary caused the RGB progenitor of the sdB star to lose
mass early on the RGB. It played the role we suggested that was played by another planet in the system V 391 Pegasi.
The M-type secondary survived the CE phase, while in V 391 Pegasi the close-in planet did not survive the CE phase.

(4)\textit{HW Vir.} Lee et al. (2009) reported the discovery of two substellar companions to HW Vir.
Similar to HS~$0705+6700$ discussed above, this system contains
an sdB star of $M_1\simeq 0.54M_\odot$ and a M-type MS star of $M_2\simeq 0.18M_\odot$ and an orbital period of
$P=2.8$h (Menzies \& Marang 1986; Kilkenny et al. 1994; Wood et al. 1993; the masses are from Drechsel et al. 2001).
The two substellar objects have the following parameters (Lee et al. 2009):
$P_1 = 15.8{\yr}$, $a_1= 5.30\pm 0.23AU$, $M_{p(1)}\sim 19.2M_J$ and $e=0.46$ for the first object, and $P_2 = 9.1{\yr}$,
$a_2=3.62 \pm 0.52AU$, $M_{p(2)}\sim 8.5M_J$ and $e=0.31$ for the second planet {{{{ (in this system the
inclination is not constrained, therefore the planet mass can be considered only as a lower limit). }}}}
Assuming that the progenitor of the sdB star had a mass of $1.25M_\odot$, and neglecting accretion by the companion,
the total initial mass is $M_{\rm Pro}=1.43 M_\odot$.
According to equation \ref{eq:radius} we find the initial orbital separations (semi-major axis)
to be $a_{0(1)} \simeq 2.7 \AU$ and $a_{0(2)} \simeq 1.8 \AU$.
Here again, the close-in M-type secondary caused the RGB progenitor to lose its mass very early on the RGB.
This enhanced mass loss process caused the planets to move outward and avoid the CE phase.
If it was not for this M-type companion, the RGB would have grown to a radius of $R_{\rm RGB} \simeq 0.5-0.7 \AU$,
before losing much mass.
The inner planet with an initial semimajor axis of $a_{0(2)} \simeq 1.8 \AU$,
and periastron distance of $(1-e)a_{0(2)} \simeq 1.3 \AU$, would have felt a very strong tidal interaction (Soker 1988a).
This tidal interaction would have caused the inner planet to spiral in.
The second planet would have survived the CE phase in that case.

As can be seen in most systems the eccentricity is relatively large (if not assumed to be zero), this
strengthens our assumption that tidal interaction is not significant.

(5) Not surprisingly, the solar system, drawn as well on Fig. \ref{fig:Multi_2_var1}, also possesses
the inner-outer planet distribution.

The main conclusions we can draw from Fig. \ref{fig:Multi_2_var1} are as follows.
In cases where a close planet that evolved through a CE phase
exist around an EHB star, it is quite likely that one or more
outer planets exist. Such planets can be detected.
In some cases the inner planet(s) will not survive the CE phase. Therefore,
we encourage a search for outer planets around EHB that have no close planet or stellar companion.
The systems HS~$0705+6700$ and HW Vir show that even when the very-close companion to the sdB star is an M-type MS star
(or even earlier MS stars), rather than a substellar object, searches for planets are still highly encourage.

If the binary hypothesis is correct, then a conclusion in the
opposite direction also holds. If an outer planet is found, most
likely a planet caused the RGB progenitor to have lost most of its
envelope. In some cases this closer-in planet (or brown dwarf)
might survive the CE phase, and be found around the EHB star.

From Fig. \ref{fig:Multi_2} we estimate that $\sim 12 - 14$
multi-planet systems will undergo a CE evolution, hence leaving a
surviving outer planet (when we set the limit as the minimum of
the bimodal distribution, i.e. $-1.5<\log
\frac{M_p}{M_J}\left({\frac{a}{AU}}\right)^2<-1$ or $-0.5<\log
\frac{M_p}{M_J}\left({\frac{a}{AU}}\right)^2<0$ respectively). If
the binary evolution hypothesis is correct, these systems will
result with EHB stars, each with an outer planet (or more). From
Fig. \ref{fig:single_2} we estimate that $\sim 70-80$ systems will
end up as red HB stars, each with at least one outer orbiting
planet (when the same limits of minimum on the bimodal
distribution are set as in Fig. \ref{fig:Multi_2}). Therefore, the
expected number according to our model of red HB stars with outer
planets (like the Sun will be) is about 5 times higher than the
number of blue HB stars with outer planets.

We note that more massive stars $M_{\rm Pro} \ga 1.5 M_\odot$ tend to have planets
mainly at large separations (Figs. \ref{fig:All_planets_3} and \ref{fig:Multi_3}; but selection effects
might be important here).
This does not influence our conclusions because such massive
stars are not likely to lose most of their envelope on the RGB due to interaction with planets.

% ==========================================================
\section{Summary and conclusions}
\label{sec:summary}
% ==========================================================

Our goal is to use the properties of known exoplanets to better
understand the role planets play in the formation of extreme
horizontal branch (EHB; sdO; sdB; hot subdwarfs) stars, and the
distribution of planets around EHB stars. EHB stars are hot HB
stars with a very low mass envelope. The explanation is that their
progenitor RGB star has lost most of its envelope on the RGB. The
key process, in cases where there is no close stellar companion, is that a
planet or a stellar companion caused this
enhanced mass loss process. We focus on the role of planets. To
lose most of its mass by interaction with a planet on the RGB the
star cannot be too massive. On the other hand the minimum mass is
determined by evolution time scale. This limits the progenitor
mass of field stars to be $M \ga 1 M_\odot$. Over all, the
relevant mass range for the main sequence (MS) progenitor is
$1M_\odot \le M_{\rm Pro} \le 1.5M_\odot$.

Following Soker \& Hershenhorn (2007), we examined the
distribution of planets according to $M_pa^2$; this is done in
Fig. \ref{fig:All_planets_2}. This figure reproduces the well
known double peak distribution. We examined the distribution for
three groups of parent star mass. From Figs.
\ref{fig:All_planets_1} and  \ref{fig:All_planets_3} it is evident
that the double peak distribution is strong only in the middle
mass range $1M_\odot \le M_{\rm Pro} \le 1.5M_\odot$
 (we note that selection effects might be important for the upper mass range).
This middle mass range coincides with that of the progenitors of EHB stars formed
by interaction with planets and brown dwarfs. We then examined
(Figs. \ref{fig:Multi_1}, \ref{fig:Multi_2} and \ref{fig:Multi_3})
the double-peak distribution for multi-planet systems. We found
that the double-peak distribution also holds for these systems in
the middle mass range.

In the binary model for the formation of EHB stars, if there is no stellar companion to the EHB star,
most likely its progenitor interacted with one or more planets or brown dwarfs.
Planets close to the progenitor, mainly in the left-peak in our figures, will enter the
CE phase at an early stage and will be destroyed (Soker 1998a).
Still, they can enhance the mass loss rate and lead to the formation of an EHB star (Soker 1998a).
Our results suggest that in many cases there are also planets in the right-peak, that can survive the
RGB evolution.
 The bimodality of planets in multiplanet system suggests that when we observe an ``outer'' planet
($a_p\geq 1AU$) around an EHB star, another close in planet was
probably engulfed during its formation process. Moreover, if we
observe a close in planet ($a\ll 1AU$) or even if we do not
observe it, planets around EHB stars are likely to reside in the
outer regions at $1 \leq a_p \leq 10AU$.

We therefore encourage a search for outer planets around EHB that have close planet as well.
Moreover, even if there is no close companion (stellar or substellar), there is a high chance
of the existence of an outer planet around the EHB star.

Furthermore, if an outer planet is found, most likely another
planet (or more) went through the CE phase and caused the RGB
progenitor to lose most of its envelope. In some cases this
closer-in planet might survive the CE phase, and be found around
the EHB star. Our general conclusion from this study is that a
single EHB (sdBO) star is likely to have an outer planet(s) in an
orbital separation of $1 \leq  a_p  \leq 10AU$. We note that red
HB stars (these are stars that maintained most of their envelope)
might also have planet at large orbital separations.

In such cases either the progenitor was too massive ($\ga 1.5
M_\odot$) for an inner planet to expel most of the progenitor
envelope, or there were no massive close in planets at all. For
example, Mercury and Venus will be engulfed when the Sun evolves
of the RGB. Earth might also be engulfed. However, these three
planets do not contain enough mass to enhance the mass loss rate
from the Sun. Therefore, in 6-7~Gyr Jupiter will orbit a red HB
star.

Comparing Fig. \ref{fig:Multi_2} and Fig. \ref{fig:single_2}
should be done with caution, taking into account that the sample
of this statistics is not large. However, it appears that
according to our model the expected number of red HB with outer
planets is $\sim 5$ times as high as that of EHB with outer
planets. Once observations will increase the number of main
sequence multi-planet systems and the number of planets around HB
stars, a population synthesis should be conducted in order to
achieve a better estimate.

These conclusions hold as well when the inner object is a low mass MS star (mainly M-type).
Indeed, HW Vir and HS 0705+6700 are such close binary systems with substellar objects around them.
We encourage the search of planets around similar binary systems, e.g., PG 1336-018
(Kilkenny et al. 1993; Drechsel et al. 2001).

We strongly suggest to look at EHB (sdO; sdB; hot subdwarfs) stars as prime targets of planet
search.

%% ==========================================================
%\title{Acknowledgments:}
%\label{sec:Acknowledgments}
% ==========================================================
 We thank an anonymous referee for comments that improved our
manuscript. This research was supported by the Asher Fund for
Space Research at the Technion, and the Israel Science foundation.
E.B. was supported in part by the Center for Absorption in
Science, Ministry of Immigrant Absorption, State of Israel.

% %%%%%%%%%%%%%%%%%%%%%%%%%%%%%%%%%%%%%%%%%%%%%%%%%%%%%%%%%%%%%%%%%%%%%%%%%%%%%%%%%%%%%
% %%%%%%%%%%%%Refrences
% %%%%%%%%%%%%%%%%%%%%%%%%%%%%%%%%%%%%%%%%%%%%%%%%%%%%%%%%%%%%%%%%%%%%%%%%%%%%%%%%%%%%%

\end{document}